\documentclass[twocolumn,floatfix,10pt]{revtex4-1}
\usepackage[utf8x]{inputenc}
\usepackage{amsmath}
\usepackage{amsfonts}
\usepackage{amssymb}
\usepackage{amsthm}
\usepackage{mathrsfs}
\usepackage{float}
\usepackage{multirow}
\usepackage{color}
\usepackage{graphics}
\usepackage{epstopdf}
\usepackage[caption=false]{subfig}
\usepackage{array}
\usepackage{subfig}
\usepackage{tikz}
\usetikzlibrary{calc}
\usetikzlibrary{decorations}
\usetikzlibrary{decorations.fractals}
\usetikzlibrary{decorations.pathreplacing}
 \usetikzlibrary{decorations.pathmorphing}
\usetikzlibrary{decorations.markings}
\usetikzlibrary{positioning,matrix}
\usetikzlibrary{shapes.arrows}
\usetikzlibrary{shapes.symbols}
\usetikzlibrary{shapes.misc}
\usetikzlibrary{shapes.geometric}
\usetikzlibrary{through}
\usetikzlibrary{mindmap,backgrounds}
\usetikzlibrary{fit}
\usetikzlibrary{arrows,positioning}
\usetikzlibrary{external}
\usepackage[colorlinks, urlcolor=black, citecolor=blue, link
color=black]{hyperref}

\def\dsp{\displaystyle}
\def \Re{\text{Re}}
\def \Im{\text{Im}}

\def\KzKzb {\ensuremath{K^0 {\kern -0.16em -\Kzb}}\xspace}

\def\PA {\ensuremath{P {\kern -0.25em A}}\xspace}

\def\Kbar {\kern 0.2em\overline{\kern -0.2em K}{}\xspace}
\def\nn    {\nonumber}
\def\sss{\scriptscriptstyle}

\def\gev{\ensuremath{\mathrm{Ge\kern -0.1em V}~}}
\def\gammaf{\ensuremath{\mathrm{\Gamma\kern -0.2em _{f}}~}}
\def\barp{{\raise.35ex\hbox{${\sss (}$}}---{\raise.35ex\hbox{${\sss )}$}}}
\def\bdbarp{\hbox{$B_d$\kern-1.4em\raise1.4ex\hbox{\barp}}}
\def\Dbarp{\hbox{$D$\kern-.85em\raise1.2ex\hbox{{{\raise.35ex\hbox{{\tiny
              (}}}--{\raise.35ex\hbox{${\sss )}$}}}}}} 
\def\deltabarp{\hbox{$\delta_0$\kern-1.08em\raise1.3ex\hbox{{{\raise.35ex
          \hbox{{\tiny (}}}--{\raise.35ex\hbox{${\sss )}$}}}}}}
\def\tev{\ensuremath{\mathrm{Te\kern -0.1em V}}}
\newcommand{\be}{\begin{equation}}
\newcommand{\ee}{\end{equation}}
\newcommand{\bea}{\begin{eqnarray}}
\newcommand{\eea}{\end{eqnarray}}

\newcommand{\modulus}[1]{\left| #1 \right|}

\allowdisplaybreaks

\begin{document}
\title{ Probing Higgs couplings at LHC and beyond}
\author{Biplob Bhattacherjee}
\email{biplobcts@iisc.ernet.in}
\affiliation{Centre for High Energy Physics Indian Institute of Science 560012
Bangalore, India }
\author{Tanmoy Modak}
\email{tanmoyy@imsc.res.in}
\author{Sunando Kumar Patra}
\email{sunandokp@imsc.res.in}
\author{Rahul Sinha}
\email{sinha@imsc.res.in}
\affiliation{The Institute of Mathematical Sciences, Taramani,
Chennai 600113, India}

\begin{abstract}
The study of the Higgs couplings following its discovery is the priority of future LHC runs.
A hint of anomalous nature will be exhibited via its coupling to the 
Standard Model(SM) particles and open up new domain of phenomenological study of physics
beyond the Standard Model. The enhanced statistics from next LHC runs  will
enable entry into the precision era to study the properties of Higgs with greater details.
In this paper we present how one can extract Higgs couplings in future LHC runs at 
$14~\tev$ via $H \rightarrow Z Z^* \rightarrow 4 \ell$, using observables constructed 
from angular distributions for the Standard Model Higgs and Higgs with mixed CP configuration. 
We show how angular asymmetries can be used to measure the ratios of the couplings and the relative phases at LHC. 
We benchmark our analysis finding out the angular asymmetries and the  best fit values of the 
ratios of the couplings for SM Higgs, CP-odd admixture, CP-even higher derivative contribution
and when CP-even higher derivative contribution and CP-odd admixture are both present. 
In the Standard Model, $HZZ$ couplings have no momentum dependence. 
It is thus essential to demonstrate the momentum independence of the couplings
to establish the couplings are SM like in nature. In this work we show how one can test the 
momentum independence of the Standard Model like coupling using angular asymmetries. 
We develop the necessary tools and demonstrate how to study the momentum dependence 
can be studied at future LHC runs.
\end{abstract}

\maketitle

\section{Introduction}\label{sec:intro} The
ATLAS~\cite{:2012gk,ATLAS:science} and CMS~\cite{:2012gu,CMS:science,:2012br}
collaboration at LHC have both discovered a new resonance of mass around
$125~\gev$ that is found to be largely consistent with the observation of a
Higgs boson.  Several studies
\cite{Nelson:1984bb,Dell'Aquila:1985ve,Nelson:1986ki,Kramer:1993jn,
Barger:1993wt, Gunion:1996xu, Miller:2001bi,Bower:2002zx,Choi:2002dq,
Choi:2002jk, Godbole:2002qu,
Buszello:2002uu,Desch:2003mw,Worek:2003zp,Kaidalov:2003fw,
Godbole:2004xe,Buszello:2006hf,Przysiezniak:2006fe,Bluj:2006,
BhupalDev:2007is,Godbole:2007cn,Godbole:2007uz,Gao:2010qx,
Englert:2010ud,Eboli:2011bq,DeSanctis:2011yc,Berge:2011ij,
Kumar:2011yta,Ellis:2012wg,Englert:2012ct, Bredenstein:2006rh,Ellis:2012jv,
Ellis:2012xd, Giardino:2012dp, Choi:2012yg, Boughezal:2012tz, Banerjee:2012ez,
Avery:2012um, Coleppa:2012eh,Geng:2012hy, Ellis:2012mj,Frank:2012wh,
Djouadi:2013-21-Jan,Englert:2012xt,Stolarski:2012ps,Gainer:2013rxa,Chen:2013waa,
Modak:2013sb,atlash2zzHL,Khachatryan:2014kca,Anderson:2013afp,Modak:2014ywa,Belyaev:2015xwa,Chen:2015iha} 
done both before and after the discovery of the Higgs boson have examined how to determine the spin,
parity and coupling of the Higgs boson. In gauge sector, decay modes such as $H\to\gamma\gamma$, 
$H\to ZZ$ and $H\to WW$ etc., where one (or both) of the $Z$'s and $W$'s are off-shell, are used to 
study the spin, parity and coupling of the Higgs boson. Observation of the decay
mode $H \to \gamma\gamma$ establishes that the discovered resonance is necessarily a boson and the
Landau-Yang theorem~\cite{Landau:1948kw, Yang:1950rg} excludes the possibility
of it having spin $J=1$. Furthermore if Higgs boson is a eigenstate of charge
conjugation, charge conjugation invariance along with observation of $H\to
\gamma\gamma$ also enforce~\cite{Barger:1993wt} that Higgs is a charge conjugation
$C=+$ state. Recent measurements~\cite{Aad:2013xqa,atlas29,cms002,atlash2zz} 
have shown that the resonance favors Spin 0 over spin 2. Moreover Ref.\cite{Khachatryan:2014kca} 
rules out pure pseudoscalar hypothesis i.e. $J^P=0^-$ at a 99.98 \% CL. 
However the discovered Higgs can still have small CP-odd admixture or higher derivative
CP-even contribution to its coupling. Angular distributions and angular asymmetries of Higgs
decay are essential to investigate whether the discovered resonance
is a CP eigenstate or a resonance with mixed CP configuration. As these angular asymmetries are 
functions of the Higgs couplings, studying them will allow us to probe the nature of the Higgs
coupling directly.

In this paper we will restrict ourselves to experimentally clean golden channel 
$H \to Z Z^*\to (\ell_1^- \ell_1^+) (\ell_2^- \ell_2^+),$ where $\ell_1$, $\ell_2$ are leptons
$e$ or $\mu$.  We consider Spin-0 Higgs boson $H$ with even parity but, include the possibility 
of a small CP-odd admixture and higher derivative CP-even contribution.
We first calculate the differential decay rate for $H \to Z Z^*\to (\ell_1^- \ell_1^+) (\ell_2^- \ell_2^+)$
process in terms of invariant mass of the dileptons coming from the off shell $Z$ boson and angular distributions
of the final state leptons. From these distributions, we construct angular asymmetries (observables) and utilize 
them to probe anomaly in $HZZ$ couplings. Similar asymmetries have also been discussed 
in Ref~.\cite{Buchalla:2013mpa, Beneke:2014sba, Modak:2014zca}.
As these observables are functions of $HZZ$ vertex factors, the values of the different observables
differ for the various cases such as SM Higgs, CP-odd admixture, CP-even higher derivative contribution. 
Ref.\cite{Modak:2013sb, Khachatryan:2014kca} have discussed how ratios of couplings can be measured  at $8$ TeV LHC run. 
Ref.\cite{Modak:2013sb} shows how using uniangular distribution as input in 
likelihood analysis one can discriminate different spin possibilities.  
In this paper a study showing how using simple angular asymmetries one can study the CP property of $H$ in future
LHC runs.

We benchmark these observables for SM Higgs and Higgs with a CP-odd admixture at $14$ 
TeV 300 fb$^{-1}$ LHC. We determine the ratios of the coupling constants and use them to discriminate possible CP-odd 
admixture from SM Higgs. We then perform the same analysis at 3000 fb$^{-1}$ to distinguish SM, case with CP-odd
admixture and CP-even higher derivative contribution. We also consider the scenario when higher derivative CP-even and 
and CP-odd admixture are present in $HZZ$ couplings. We denote this scenario as `CP-even-odd' case.
In our analysis we have also included a complex phase for CP-odd admixture Higgs and show how to determine the phase using angular
asymmetries. Furthermore we use these angular asymmetries and perform chi-square analysis to probe both CP-even and CP-odd
anomalous contributions in the Higgs couplings.

Exotic models of Higgs can have momentum dependence in its couplings. It is thus essential
to study the momentum dependence of the Higgs couplings to establish its SM nature. Angular asymmetries will provide
necessary tools to investigate the momentum dependence of the Higgs couplings. In our work we develop necessary techniques
and utilize them to probe the momentum dependence of $HZZ$ couplings. It is important to note this it will definitely 
require higher statistics. However the enhanced statistics at $14$ TeV, LHC will enable us to study the 
momentum dependence of Higgs couplings 
in different momentum regions. Since the mass of $H$ does not allow both the $Z$ bosons to be on-shell, the invariant mass 
distribution of the dileptons from the off-shell $Z$ boson($Z_2$) will offer us a test for the momentum dependence
of the $HZZ$ couplings. The $HZZ$ couplings in the most general case could be function of the invariant mass of 
the off-shell $Z$. The constancy of the ratios of the Higgs couplings can be measured by finding out the values of the 
ratios of couplings in different momentum regions for the invariant mass of the off-shell $Z$. In our analysis we have 
shown for SM how one can test the momentum dependence of the Higgs couplings at 14 TeV $300$ fb$^{-1}$ LHC run.

In LHC we measure $\sigma \cdot BR$ and that does not allow us to measure
the decay width of the Higgs, as a result we can only measure the ratios of Higgs couplings. However 
ILC\cite{Baer:2013cma} will be able to measure the inclusive cross section($\sigma_{ZH}$) using recoil technique for the process
$e^+ e^- \rightarrow Z H$. Hence inclusive cross section provide a direct measurement of $HZZ$ couplings at ILC by measuring
partial width $\Gamma(H\rightarrow Z Z)$. After energy upgrade LHC will run at $33$ TeV and enhanced statistics
will enable us to measure Higgs couplings more precisely than that possible at $14$ TeV.
We discuss how much this energy upgrade will improve the measurement of $HZZ$ couplings and  
discuss what sensitivity can be achieved compared to that of an ILC\cite{Asner:2013psa} measurement.
 
The paper is organized as follows:
In Section~\ref{sec:form} we formalize the necessary tools for our analysis. Section~\ref{sec:num} is divided into
three subsection. First we benchmark our analysis for 300 fb$^{-1}$ in Subsection~\ref{subsec:higgs300} and then
we benchmark our analysis for 3000 fb$^{-1}$ for SM Higgs, Higgs with CP-odd admixture, CP-even higher derivative 
contribution and CP-even-odd scenario in Subsection~\ref{subsec:higgs3000}. 
In Subsection~\ref{subsec:madep} we develop the necessary technique to measure the momentum dependence of Higgs couplings. 
A qualitative comparison between future runs of LHC and ILC precision measurement has been made in Section~\ref{comp}. 
Finally we conclude in Section~\ref{sec:conclusion}.
\section{The Formalism}
In this section we first write down the $HZZ$ vertex, helicity amplitudes in transversity basis and 
finally derive the expression for angular distribution of $H \to Z Z^* \to 4 \ell$ process assuming $H$ 
to be a spin $0$ particle. In SM the process $H \to Z Z$  is characterised by the Lagrangian
\begin{align}
 \mathcal{L}_{HZZ} &= \frac{gM_Z}{2\cos\theta_W}\,  Z_\mu Z^\mu H \label{eq:sm}
\end{align}
where $\theta_W$ is the Weinberg angle and $g$ is the electroweak coupling constant.
However there may exist anomalous couplings of $H$ to $Z$ boson. These couplings can in general
be CP-even or CP-odd and can be generated from the effective Lagrangians
\begin{align}
\mathcal{L}_{e} &\sim- \frac{1}{4} Z^{\mu \nu}Z_{\mu \nu} \,H
\end{align} and  
\begin{align} 
\mathcal{L}_{o} &\sim- \frac{1}{4} Z^{\mu \nu}\tilde{Z}_{\mu \nu}\,H 
\end{align} respectively, where
$Z_{\mu\nu}$  and $\tilde{Z}_{\mu\nu}$ are defined  as
$Z_{\mu\nu} = \partial_{\mu} Z_{\nu} - \partial_{\nu} Z_{\mu}$ and 
$\tilde{Z}_{\mu\nu} = \epsilon_{\mu\nu\rho\sigma} \, Z^{\rho\sigma}$ respectively.

\begin{figure}[htb!]
\centering
\includegraphics[width=0.48 \textwidth]{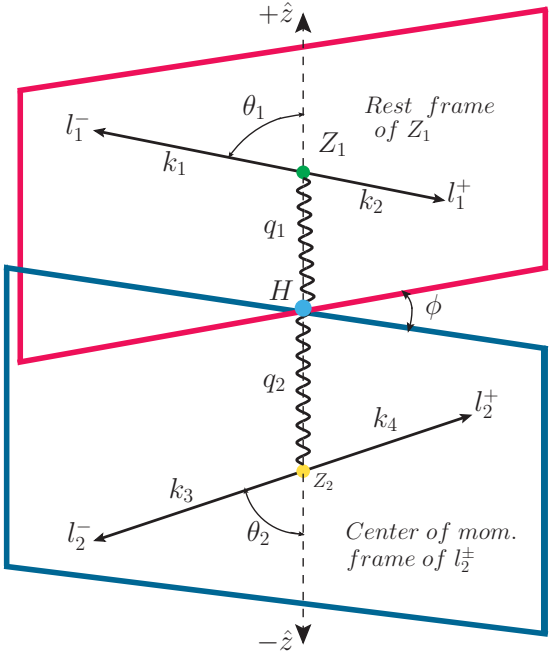}
\caption{The definition of the polar angles ($\theta_1$ and $\theta_2$)
  and the azimuthal angle ($\phi$) in the decay of Higgs ($H$) to a
  pair of $Z$ boson, followed by a decay of both the $Z$  into four charged 
  leptons: $H\to Z_1 + Z_2 \to
  (\ell_1^- + \ell_1^+) + (\ell_2^- + \ell_2^+),$ where $\ell_1,
  \ell_2 \in \{ e,\mu \}$ and three momentum
  $\vec{k}_1 = -\vec{k}_2$ and $\vec{k}_3 = -\vec{k}_4$.The lepton pair
  $\ell_1^\pm$ going back to back in the rest frame of $Z_1$, however 
  lepton pair from $Z_2$ going back to back in 
  their center-of-momentum (C.O.M) frame.}
\label{fig:HZZst}
\end{figure}

\label{sec:form}
Following these Lagrangians one can write down the most general $HZZ$ vertex as follows   
\begin{align}\label{eq:vertex}
V^{\mu\nu}&=\displaystyle\frac{igM_Z}{\cos\theta_W} 
\bigg( a \, g^{\mu\nu} + b \, \left(q_2^{\mu}q_1^{\nu} - q_1\cdot q_2\;g^{\mu \nu}\right)\nn\\ 
&  \quad \quad \quad \quad \quad \quad + i c \,\epsilon^{\mu\nu\rho\sigma} \; q_{1\rho}\,q_{2\sigma} \bigg), 
\end{align}
where  $a$, $b$, $c$ are momentum dependent vertex factors and $q_1$, $q_2$ and $P$ are the
four momenta of $Z$($Z_1$), $Z^*$($Z_2$) and $H$ respectively. Off-shellness of the $Z$ is denoted 
by the superscript `*'. In Standard Model at tree level the values of the vertex factors 
are $a=1$ and $b=c=0$ and they are constant. However a non zero $b$ and $c$ can arise from higher order
correction. If the vertex factors $a$, $b$, $c$ show any deviations from SM values or exhibit 
a momentum dependence it would provide a hint about the non standard nature of the $HZZ$ couplings. The CP-ood
admixture is charactersided by the non zero value of $c$ of the form $c\:e^{i \delta}$, where $\delta$
is the CP violating phase associated with $c$.

The decay under consideration can be characterised by three helicity amplitudes $\mathcal{A}_L$, $\mathcal{A}_{\parallel}$ and $\mathcal{A}_{\perp}$ defined as (see appendix) :
\begin{align}
  \mathcal{A}_L &= q_1 \cdot q_2 \,\left( a - b~q_1.q_2 \right) + M_H^2\, X^2 \, b, \label{eq:AL}\\
 \mathcal{A}_{\parallel} &= \sqrt{2 q_1^2 \, q_2^2} \, \left( a - b ~q_1.q_2 \right), \label{eq:AA}\\
 \mathcal{A}_{\perp} &= \sqrt{2 q_1^2\, q_2^2}\, X \, M_H \, c, \label{eq:AP}
\end{align}
where $\sqrt{q_1^2}$ and $\sqrt{q_2^2}$ are the invariant masses of
the $Z_1$ and $Z_2$ respectively, with
\begin{equation}\label{eq:x}
  X=\dsp\frac{\sqrt{\dsp\lambda(M_H^2,q_1^2,q_2^{2})}}{\dsp 2M_H},
\end{equation} and
\begin{equation}
  \lambda(x,y,z)= x^2+y^2+z^2-2\,x\,y-2\,x\,z-2\,y\,z~.
\end{equation} 
The helicity amplitudes have definite parity properties as $\mathcal{A}_L$, $\mathcal{A}_{\parallel}$ are CP-even and $\mathcal{A}_{\perp}$ is CP-odd.

Having defined the helicity amplitudes in transversity basis the full angular distribution for 
$H\to Z_1 + Z_2 \to (\ell_1^- + \ell_1^+) + (\ell_2^- + \ell_2^+),$ can be written as\cite{Modak:2013sb}
\begin{widetext}
\begin{align}\label{eq:fangdist}
  &\frac{8\pi}{\gammaf}\frac{d^4\Gamma}{dq_2^2\; d\cos{\theta_1} \;
    d\cos{\theta_2} \; d\phi} = 1 +\frac{|\mathcal{F}_\||^2-|\mathcal{F}_\perp|^2}{4}
  \cos\,2\phi\big(1-P_2(\cos\theta_1)\big) 
  \big(1-P_2(\cos\theta_2)\big)+\frac{1}{2} \Im(\mathcal{F}_{\parallel} \mathcal{F}_{\perp}^*)\,\sin\,2\phi\nn\\
  &\times\big(1-P_2(\cos\theta_1)\big)\big(1-P_2(\cos\theta_2)\big)
  +\frac{1}{2}(1-3\modulus{\mathcal{F}_L}^2)\,\big(P_2(\cos\theta_1)+P_2(\cos\theta_2)\big) +
  \frac{1}{4}(1+3\modulus{\mathcal{F}_L}^2)\,P_2(\cos\theta_1)P_2(\cos\theta_2)\nn\\
  &+\frac{9}{8\sqrt{2}} \left( \Re(\mathcal{F}_L \mathcal{F}_{\parallel}^*)\,\cos\phi + \Im(\mathcal{F}_L \mathcal{F}_{\perp}^*)\,\sin\phi\right)
  \sin\,2\theta_1\,\sin\,2\theta_2\; +\eta\frac{9}{2\sqrt{2}}\Re(\mathcal{F}_L \mathcal{F}_{\perp}^*)
  \big(\cos\theta_1-\cos\theta_2)\cos\phi \sin\theta_1\sin\theta_2\nn\\
 & +\eta\frac{3}{2} \Re(\mathcal{F}_{\parallel}\mathcal{F}_{\perp}^*)\big(\cos\theta_2 (2 + P_2(\cos\theta_1))
 - \cos\theta_1(2 + P_2(\cos\theta_2))\big)-\eta\frac{9}{2\sqrt{2}}\Im(\mathcal{F}_L \mathcal{F}_{\parallel}^*)
  \big(\cos\theta_1-\cos\theta_2)\sin\phi \sin\theta_1\sin\theta_2\nn\\
  &-\frac{9}{4}\eta^2\Bigg((1-\modulus{\mathcal{F}_L}^2)\cos\theta_1\cos\theta_2
  + \sqrt{2} \left( \Re(\mathcal{F}_L \mathcal{F}_{\parallel}^*)\cos\phi + \Im(\mathcal{F}_L
    \mathcal{F}_{\perp}^*)\sin\phi \right) \sin\theta_1\sin\theta_2\Bigg),
\end{align}
\end{widetext}
where  $\mathcal{F}_L$, $\mathcal{F}_{\parallel}$,
$\mathcal{F}_{\perp}$ are the {\it{helicity fractions}} defined in the appendix along with $\gammaf$ and $\eta$. The angle $\theta_1$($\theta_2$) is the angle between three momenta of $\ell_1^-$ ($\ell_2^-$) in $Z_1$($Z_2$) rest frame and the direction of three momenta of $Z_1$($Z_2$) in $H$ rest frame. The angle $\phi$ is defined as the angle between the normals to the planes defined by  $Z_1\to \ell_1^- \ell_1^+$ and  $Z_2\to \ell_2^- \ell_2^+$ in $H$ rest frame as shown in
Fig.~\ref{fig:HZZst}. It should be noted that Eq.(\ref{eq:fangdist}) is \textit{exact} and \textit{ 
no assumptions has been made apart from assuming that the leptons are massless}.

Integrating Eq.(\ref{eq:fangdist}) with respect to any two out of the three angles $\theta_1$, $\theta_2$, $\phi$
one finds three uniangular distributions for the process $H \to ZZ\to (\ell_1^- \ell_1^+) (\ell_2^- \ell_2^+)$
as:
\begin{align}
  \frac{1}{\gammaf}\frac{d^2\Gamma}{dq_2^2 \; d\cos\theta_1} &=
  \frac{1}{2} + T_2\,P_2(\cos\theta_1) - T_1
  \cos\theta_1, \label{eq:cost1} \\
  \frac{1}{\gammaf}\frac{d^2\Gamma}{dq_2^2 \; d\cos\theta_2} &=
  \frac{1}{2} + T_2\,P_2(\cos\theta_2) + T_1
  \cos\theta_2, \label{eq:cost2} \\
  \frac{2\pi}{\gammaf}\frac{d^2\Gamma}{dq_2^2 \; d\phi} &= 1 +
  U_2\,\cos\,2\phi + V_2\,\sin\,2\phi \nn\\
 & + U_1\cos\phi + V_1 \sin\phi, \label{eq:phi}
\end{align}

The observables  $T_1$, $T_2$, $U_1$, $U_2$, $V_1$ and $V_2$ are defined as:
\begin{align}
\label{eq:T10}
T_1 &=\frac{3}{2} \, \eta \Re(\mathcal{F}_{\parallel} \mathcal{F}_{\perp}^*),\\
T_2 &=\frac{1}{4} (1-3\modulus{\mathcal{F}_L}^2),\\
U_1 &=-\frac{9\pi^2}{32\sqrt{2}} \eta^2 \, \Re(\mathcal{F}_L \mathcal{F}_{\parallel}^*),\\
U_2 &=\frac{1}{4} (|\mathcal{F}_{\parallel}|^2-|\mathcal{F}_{\perp}|^2),\\
V_1 &=-\frac{9\pi^2}{32\sqrt{2}} \eta^2 \, \Im(\mathcal{F}_L \mathcal{F}_{\perp}^*),\\
V_2 &=\frac{1}{2}\,\mathrm{Im}(\mathcal{F}_{\parallel}\mathcal{F}_{\perp}^*),
\end{align}
and are functions of $q_2^2$. Moreover  $P_1(\cos\theta_{1,2})$  $P_2(\cos\theta_{1,2})$ are 
first and second degree Legendre Polynomials respectively. It should be noted that the observables
$T_1$, $T_2$, $U_1$, $U_2$, $V_1$ and $V_2$ are coefficients of orthogonal functions $P_2(\cos\theta_{1,2})$, 
$P_1(\cos\theta_{1,2})$, $\cos2\phi$, $\cos \phi$, $\sin2\phi$, $\sin\phi$ respectively and can be extracted
individually. These observables are functions of the helicity fractions $\mathcal{F}_L$, $\mathcal{F}_{\parallel}$ and
$\mathcal{F}_{\perp}$, hence they are functions of vertex factors $a$, $b$ and $c$. Measurement of these
observables will enable us to probe the vertex factors $a$, $b$ and $c$.
Moreover in SM  $T_2$,  $U_2$ and $U_1$ all are non zero as $\mathcal{F}_L$ and $\mathcal{F}_{\parallel}$ 
are non zero. In SM at tree level $c=0$ which enforces $\mathcal{F}_{\perp}=0$. 
As $T_1$, $V_2$ and $V_1$ all are functions of $\mathcal{F}_{\perp}=0$, in SM they are all zero. 
A CP-odd admixture is characterised by non zero value of $c$ and hence non vanishing values of
$T_1$, $V_2$ and $V_1$. Measurements of $T_1$, $V_2$ and $V_1$ allow us to probe CP violating phase in
$HZZ$ couplings. If there exist any CP-even higher derivative contribution in $HZZ$ couplings, the vertex factor
$b$ becomes non zero, hence the observables $T_2$, $U_2$ and $U_1$ will have different values than that of SM.
In SM at tree level $b=0$ but at one loop level the value of $b$ will be non zero. Measurement of $b$ 
will allow us to probe triple-Higgs vertex which arises at one loop level and provide the first 
verification of Higgs self coupling.

This observables can be extracted using following angular asymmetries 
\begin{widetext}
\begin{align}
  T_1 &= \left( \int_{-1}^{0} - \int_{0}^{+1} \right)
  d\cos\theta_1 \; \left( \frac{1}{\gammaf}\frac{d^2\Gamma}{dq_2^2 \;
    d\cos\theta_1} \right) = \left( -\int_{-1}^{0} + \int_{0}^{+1}
  \right) d\cos\theta_2 \; \left( \frac{1}{\gammaf}
  \frac{d^2\Gamma}{dq_2^2 \; d\cos\theta_2} \right), \label{eq:T10asym}\\%
  T_2 &= \frac{4}{3} \left( \int_{-1}^{-\frac{1}{2}} -
  \int_{-\frac{1}{2}}^{+\frac{1}{2}} + \int_{+\frac{1}{2}}^{+1}
  \right) d\cos\theta_{1,2} \; \left( \frac{1}{\gammaf}
  \frac{d^2\Gamma}{dq_2^2 \; d\cos\theta_{1,2}} \right),\label{eq:T20asym}\\%
  U_1 &= \frac{1}{4} \left( -\int_{-\pi}^{-\frac{\pi}{2}} +
  \int_{-\frac{\pi}{2}}^{+\frac{\pi}{2}} -
  \int_{+\frac{\pi}{2}}^{+\pi} \right) d\phi \; \left(
  \frac{2\pi}{\gammaf} \frac{d^2\Gamma}{dq_2^2 \; d\phi}
  \right), \label{eq:U10asym}\\%
  U_2 &= \frac{\pi}{2 \gammaf} \left( \int_{-\pi}^{-\frac{3\pi}{4}} -
  \int_{-\frac{3\pi}{4}}^{-\frac{\pi}{4}} +
  \int_{-\frac{\pi}{4}}^{\frac{\pi}{4}}
  -\int_{\frac{\pi}{4}}^{\frac{3\pi}{4}} + \int_{\frac{3\pi}{4}}^{\pi}
  \right) d\phi \; \frac{d^2\Gamma}{dq_2^2\,d\phi} ,\label{eq:U20asym}\\%
  V_1 &= \frac{1}{4} \left( - \int_{-\pi}^{0} + \int_{0}^{+\pi}
  \right) d\phi \; \left( \frac{2\pi}{\gammaf}\frac{d^2\Gamma}{dq_2^2
    \; d\phi} \right),\label{eq:V10asym}\\%
  V_2 &= \frac{1}{4} \left( \int_{-\pi}^{-\frac{\pi}{2}} -
  \int_{-\frac{\pi}{2}}^{0} + \int_{0}^{+\frac{\pi}{2}} -
  \int_{+\frac{\pi}{2}}^{+\pi} \right) d\phi \; \left(
  \frac{2\pi}{\gammaf}\frac{d^2\Gamma}{dq_2^2 \; d\phi}
  \right).\label{eq:V20asym}%
\end{align}
\end{widetext}

Three  uniangular distributions in Eq.(\ref{eq:cost1}), Eq.(\ref{eq:cost2}) and Eq.(\ref{eq:phi}) give 6 observables 
$T_1$, $T_2$, $U_1$, $V_1$ and $V_2$ in terms of angular asymmetries shown above.  At 14 TeV, high luminosity future LHC runs will 
provide enhanced statistics and the uniangular distributions will not only allow us to probe Higgs couplings, 
but also provide us with arsenal to probe the momentum dependence of the Higgs couplings which is 
essential for precision measurement. In the next section we will discuss these aspects in detail.
\section{Numerical Study}
\label{sec:num}
In this section we demonstrate how angular analysis can be used to benchmark different CP scenarios of $H$.
We have generated events with MADEVENT5~\cite{Alwall:2011uj} event generator
interfaced with PYTHIA6.4~\cite{Sjostrand:2003wg} and Delphes 3~\cite{deFavereau:2013fsa}. 
The vertex, Eq.~(\ref{eq:vertex}) is parametrized by UFO format of MadGraph5 using HiggsCharacterisation
model~\cite{Artoisenet:2013puc}. The events are generated by $pp$ collisions via $g g \to H$ and $ g g \to H + 1 jet$,
for center of mass energy $\sqrt{s}=14$ TeV, using parton distribution functions CTEQ6L1~\cite{Pumplin:2002vw}.

For matching purpose we have used MLM prescription and events are finally passed through 
fast detector simulator package Delphes 3. We are only concerned about the decay process of the
$H$ and have assumed only the SM production process of $H$ while generating events. 

We follow the cut based analysis of Ref.\cite{atlash2zz}. The identical pairs of final 
state leptons $2 e^+ 2 e^-$ and $2\mu^+ 2 \mu-$ events have also been taken into account
for our analyses. The on-shell $Z_1$ boson is identified by the invariant mass of the opposite
charge same flavor lepton pairs closest to $M_Z$. Moreover, since the Higgs mass $M_H$ does not allow both
the $Z$ bosons to be on-shell, this in turn breaks the need for Fermi antisymmetrisation of the 
identical fermions when the final states are $2 e^+ 2 e^-$ and $2\mu^+ 2 \mu-$. The branching ratios 
and cross sections have been taken from Higgs working Group webpage~\cite{hwgpage}.

Following the analysis presented in Ref.\cite{atlash2zz} data are selected using single-lepton or 
di-lepton triggers. For the single-muon trigger the transverse momentum, $P_T$, threshold is 25 GeV, 
while for single-electron trigger the transverse energy, $E_T$, threshold is 25 GeV. Di-muon 
triggers are selected using two ways. For asymmetric di-muon the trigger thresholds are either
$p_{T1} = 18$ GeV and $p_{T2} = 8$ GeV. Threshold for symmetric di-muon triggers  are $p_T = 13$ GeV for both the muons.
For the di-electron trigger the thresholds are $E_T = 12$ GeV for both electrons. There are two electron-muon 
triggers used with 12 or 24 GeV $E_T$ electron thresholds, differed by the electron identification requirements, 
and muon threshold $p_T=8$ GeV.

Each electron (muon) must satisfy $E_T>$ 7 GeV ($p_T >$ 6 GeV) and be measured in the pseudo-rapidity
range $|\eta|<$   2.47 ($|\eta| <$ 2.7). 
We have selected the leptons in two sequential $p_T$ ordered way.\\
i) Case-I : $p_T$  of at least   two leptons in a quadruplet must satisfy $p_T >$ 20 GeV,\\
ii) Case-II :  $p_T$  of at least  three leptons in a quadruplet must satisfy $p_T >$ 20 GeV.\\ 
 \\

The leptons are required to
be separated from each other by $\Delta R>$ 0.1 if they are of the same flavour and  $\Delta R >$ 0.2 otherwise. Each
event is required to have the triggering lepton(s) correctly matched to one or two of the selected leptons.

Furthermore we also impose the invariant mass cuts on the
$m_{Z_1}(\sqrt{q_1^2})$, $m_{Z_2}(\sqrt{q_2^2})$ and $m_{4\ell}$ described in
Table~\ref{cut_table}. $m_{Z_1}$ is the invariant mass of the pair of opposite
sign same flavor leptons closest to $m_Z$ while $m_{Z_2}$ is the other
combination. The two columns of Table~\ref{cut_table} demonstrate
the effect of $p_T$ ordering in event selection.
\begin{table}[hbtp!]
\centering
\begin{tabular}{|c|c|c|}
\hline
Cuts & Case-I  & Case-II\\
\hline
Selection  cuts                           & 494     &  2253 \\
$50\mbox{ GeV} <m_{12} < 106$~GeV         & 487     &  2204  \\
$12\mbox{ GeV} <m_{34} < 115$~GeV         & 447     &  2071  \\
$115 \mbox{ GeV} < m_{4\ell} < 130$~GeV   & 443     &  2050  \\
\hline
\end{tabular}
\caption{Effects of the sequential cuts on the simulated Signal for two different $p_T$ ordering of Case-I(first column) 
and Case-II( second column). The sequential $p_T$ ordering of Case-I is for 300 fb$^{-1}$, however
we have used sequential $p_T$ ordering of Case-II for 3000 fb$^{-1}$.
The $k$-factor for signal is $2.5$.}
\label{cut_table}
\end{table}

Now integrating Eq.~\eqref{eq:cost1},\eqref{eq:cost2}, \eqref{eq:phi} 
over $q_2^2$ we get three integrated distributions as follows
\begin{align}
\frac{1}{\Gamma}\frac{d\Gamma}{d\cos\theta_1} &= \frac{1}{2} -
\mathcal{T}_1 \, \cos\theta_1 +
\mathcal{T}_2 \, P_2(\cos\theta_1),\label{eq:qct1} \\
\frac{1}{\Gamma}\frac{d\Gamma}{d\cos\theta_2} &= \frac{1}{2} +
\mathcal{T}_1 \, \cos\theta_2 +
\mathcal{T}_2 \, P_2(\cos\theta_2),\label{eq:qct2}\\
\frac{1}{\Gamma}\frac{d\Gamma}{d\phi} &= \frac{1}{2\pi} +
\mathcal{U}_1 \, \cos\phi + \mathcal{U}_2
\,\cos2\phi \nn\\
& \quad  \quad \quad+\mathcal{V}_1 \, \sin\phi + \mathcal{V}_2
\,\sin2\phi,\label{eq:qphi}
\end{align} 
where $\mathcal{T}_1$, $\mathcal{T}_2$, $\mathcal{U}_1$, $\mathcal{U}_2$
,  $\mathcal{V}_1$ and  $\mathcal{V}_2$ are observables integrated over $m_{34}$($q_2^2$) and $m_{12}$.

\begin{figure}[hbtp!]
\centering
\includegraphics[width=0.69\linewidth, angle =-90]{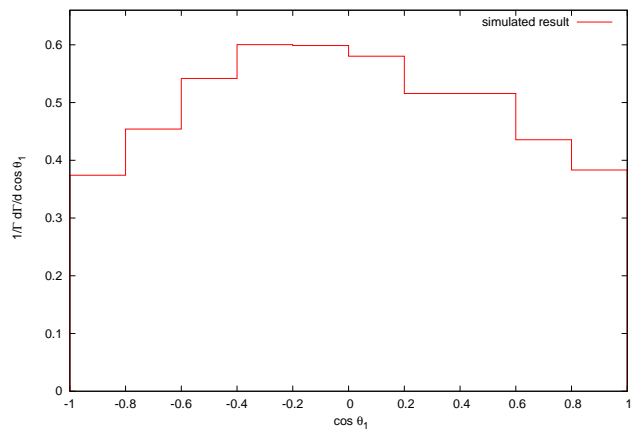} 
\caption{The normalized  distribution $\frac{1}{\Gamma} \frac{d\Gamma}{d\cos\theta_1}$ vs $\cos\theta_1$ for SM Higgs events.}
\label{fig:ct1dist}
\end{figure}

\begin{figure}[hbtp!]
\centering
\includegraphics[width=0.69\linewidth, angle =-90]{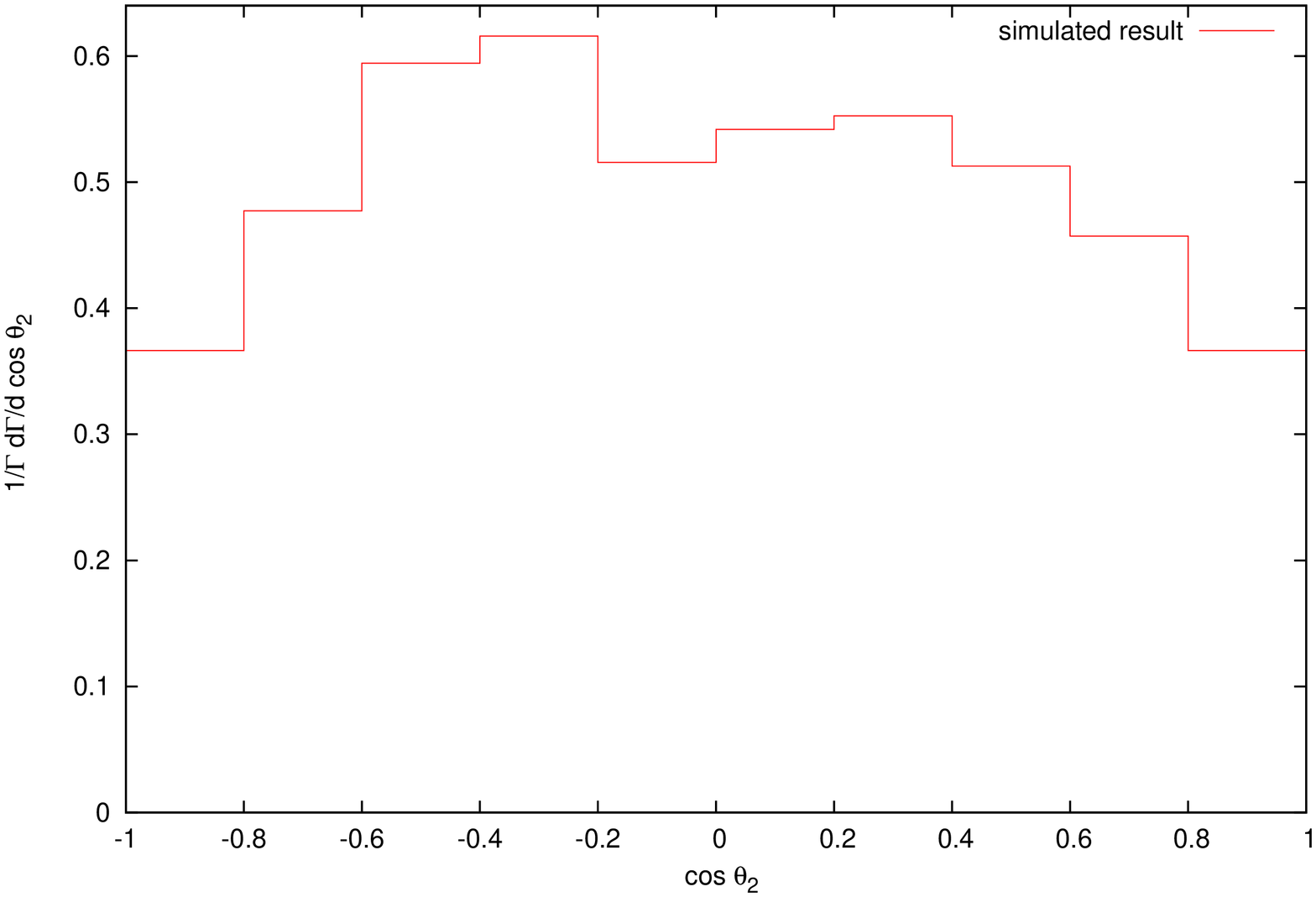} 
\caption{The normalized  distribution $\frac{1}{\Gamma} \frac{d\Gamma}{d\cos\theta_2}$ vs $\cos\theta_2$ for SM Higgs events. }
\label{fig:ct2dist}
\end{figure}

\begin{figure}[hbtp!]
\centering
\includegraphics[width=0.69\linewidth, angle =-90]{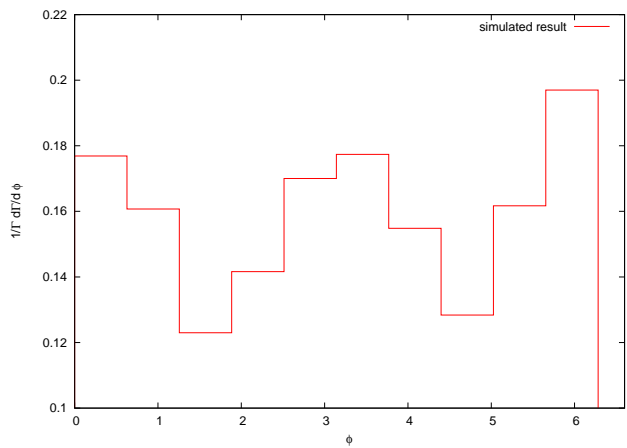} 
\caption{The normalized  distribution $\frac{1}{\Gamma} \frac{d\Gamma}{d\phi}$ vs $\phi$ for SM Higgs events. }
\label{fig:phidist}
\end{figure}

The normalized distributions,  $\frac{1}{\Gamma} \frac{d\Gamma}{d\cos\theta_1}$ vs $\cos\theta_1$ 
,  $\frac{1}{\Gamma} \frac{d\Gamma}{d\cos\theta_2}$ vs $\cos\theta_2$  and  
$\frac{1}{\Gamma} \frac{d\Gamma}{d\phi}$ vs $\phi$ for
SM are shown in Fig.~\ref{fig:ct1dist}, Fig.~\ref{fig:ct2dist} and Fig.~\ref{fig:phidist} respectively 
for simulated data. It should be noted that the angular coverage for
$\cos\theta_1$ or $\cos\theta_2$ covers the full range from $-1$ to $+1$
and coverage for $\phi$ from $0$ to $2\pi$ are still retained even after using actual detector scenarios. 
The cut flow analysis of Case-I is followed for the  analysis of SM Higgs and Higgs with CP-odd admixture at 300 fb$^{-1}$. 
At 3000 fb$^{-1}$ since the statistics is higher, we will use stronger cut based analysis i.e. sequential cut flow 
analysis of Case-II for benchmarking SM Higgs and Higgs with different CP configuration. 
Moreover it should be noted that we have used the same cut based analysis for CP-odd admixture, 
CP-even higher derivative contribution and CP-even-odd scenario.
The cross section for each benchmark scenarios are within the current experimental allowed region.


The simulated data are binned in $\cos\theta_1$,  $\cos\theta_2$ and $\phi$
and fitted using Eq.\eqref{eq:qct1}, Eq.\eqref{eq:qct2} and Eq.\eqref{eq:qphi}  
to obtain the angular asymmetries $t_1$, $t_2$, $u_1$, $u_2$, $v_1$, $v_2$ and their
errors which correspond to the angular asymmetries $\mathcal{T}_1$, $\mathcal{T}_2$, $\mathcal{U}_1$, $\mathcal{U}_2$
, $\mathcal{V}_1$ and  $\mathcal{V}_2$ respectively.
 Once the values of the integrated observables $t_1$, $t_2$, $u_1$, $u_2$, $v_1$, $v_2$ 
and their respective errors are found, the $\chi^2$ formula:
\begin{align}
 \chi^2 &=\frac{(\mathcal{T}_2-t_2)^2}{(\Delta t_2)^2}+\frac{(\cos\delta \; \mathcal{T}_1 -t_1)^2}{(\Delta t_1)^2}
    +\frac{(\mathcal{U}_1-u_1)^2}{(\Delta u_1)^2} \nn\\ 
     &+\frac{(\mathcal{U}_2-u_2)^2}{(\Delta u_2)^2}
    +\frac{(\sin\delta\;\mathcal{V}_2-v_2)^2}{(\Delta {v}_2)^2}
    +\frac{(\sin\delta\;\mathcal{V}_1-v_1)^2}{(\Delta {v}_1)^2}\label{eq:chisq}
\end{align}
will find the $b/a$, $c/a$ and the phase $\delta$. The errors in $b/a$, $c/a$ and phase $\delta$ can also be
calculated using the error matrix 
\begin{equation}
 \left(\frac{\partial^2\chi^2}{\partial \alpha_i \partial \alpha_i}\right)_{\hat{\alpha}}
\end{equation}
where  $\alpha_i,\alpha_j = b/a, c/a, \delta$. To find the best fit values we have used Mathematica 9\cite{mmatica}.

\subsection{Study of  angular asymmetries of Higgs at 14 TeV and 300 fb$^{-1}$}
\label{subsec:higgs300}

We start benchmarking angular observables for SM Higgs and Higgs with CP-odd admixture in this section. 
The fit values for observables are tabulated along with the best fit values of $b/a$, $c/a$ and phase $\delta$ of 
the CP-odd coupling. We also obtain $1\sigma$ and $2\sigma$ contours for $b/a$ vs $c/a$ and 
$\delta$ vs $c/a$ for the case of CP-odd admixture. This will provide the precision at which one can rule 
out anomalous contributions in $HZZ$ couplings, establishing the SM nature of $H$ at 14 TeV 300 fb$^{-1}$ LHC run.

\subsubsection{SM Higgs}
\label{sm}

SM Higgs events are generated with $a=1$, $b=0$ , $c=0$. The fit values of the observables for the SM Higgs are 
tabulated in Table~\ref{smhiggs}.

\begin{table}[hbtp!]
\caption{The values of the observables for the SM Higgs with respective errors.}
\centering
\begin{tabular}{c |c }
\hline
\hline
\textbf{Observables} &    \textbf{Values with errors}  \\
\hline

$t_2$  &     $-0.21 \pm 0.09$                         \\

$t_1$  &     $(-1.6 \pm 7.16)\times10^{-2}$         \\

$u_2$  &     $0.32 \pm 0.40$        \\

$u_1$  &      $(0.93\pm 4.36 )\times10^{-1}$        \\

$v_2$  &     $(-0.72\pm 4.03 )\times10^{-1}$        \\

$v_1$  &     $(0.19\pm 3.70 )\times10^{-1}$         \\
\hline
\end{tabular}
\label{smhiggs}
\end{table}
The values of the observables $t_2$ and $u_2$ are large compared to other observables as discussed 
in the previous section, playing important role in the $\chi^2$ expression in Eq.\eqref{eq:chisq}. 
The observables $t_1$, $v_2$ and $v_1$ provide information about phase for anomalous couplings
$b$ and $c$. The best fit values of $b/a$ and $c/a$  with their respective errors for the SM Higgs
are given as follows:
\begin{align}
 b/a &= (0.50 \pm 0.96)\times10^{-4}\,\mbox{GeV}^{-2} \\
 c/a &= (0.68\pm 2.27)\times10^{-4}\,\mbox{GeV}^{-2}
\end{align}

\begin{figure}[hbtp!]
\centering
\includegraphics[width=0.32\textwidth]{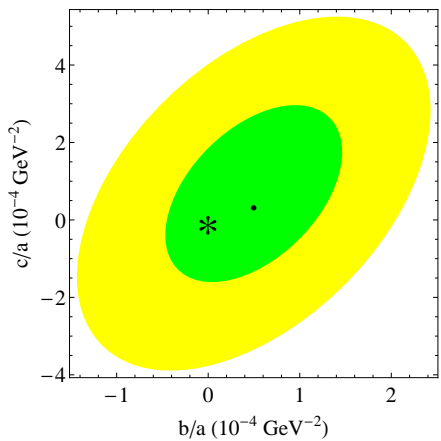} 
\caption{$c/a$ vs $b/a$ $1\sigma$ (green) and $2\sigma$ (yellow)
  contours for the SM Higgs at 300 fb$^{-1}$.  The best fit values $(b/a, c/a)$ is
  shown by the block dot. The $`*$' corresponds to $b = c = 0$.}
\label{fig:bacasm}
\end{figure}

The best fit values with $1\sigma$ and $2\sigma$ contours for
$b/a$ vs $c/a$ are shown in Fig~.\ref{fig:bacasm}.

\subsubsection{Higgs with CP-odd admixture}
The CP-odd admixture is charaterised by a non zero value of $c$ in Eq.(\ref{eq:vertex}). 
For CP-odd admixture case, Higgs events are generated using $a=0.7$, $b=0$ and $c=(2.2+2.2 i)\times10^{-4}$.
The values of the observables are given in Table~\ref{oddadmix-300}. 

\begin{table}[hbtp!]
\caption{The values of the observables for CP-odd admixture Higgs with respective errors at 300 fb$^{-1}$.}
\centering
\begin{tabular}{c|c }
\hline
\hline
\textbf{Observables} &    \textbf{Values with errors}  \\
\hline

$t_2$  &     $-0.06 \pm 0.10$                         \\

$t_1$  &     $-0.11 \pm 0.08$         \\

$u_2$  &     $-0.08 \pm 0.40$        \\

$u_1$  &      $(-0.80\pm 4.04 )\times10^{-1}$        \\

$v_2$  &     $(0.99\pm 4.20 )\times10^{-1}$        \\

$v_1$  &     $(0.44\pm 4.21 )\times10^{-1}$         \\
\hline
\end{tabular}
\label{oddadmix-300}
\end{table}

The value of $t_2$ has now become smaller compared to the SM case as shown in Table~\ref{smhiggs}. 
Most importantly the non zero value of $t_1$ arises due to the complex CP-odd anomalous coupling $c$. 
This will play a significant role along with $t_2$ and $u_2$ in probing anomalous CP-odd admixture of $HZZ$ couplings.
The best fit values for $b/a$, $c/a$ and the phase $\delta$ for CP-odd admixture are : 
\begin{align}
 b/a &= (1.50 \pm 1.09)\times10^{-4}\,\mbox{GeV}^{-2} \\
 c/a &= (5.48\pm 1.12)\times10^{-4}\,\mbox{GeV}^{-2} \\
 \delta & = (0.29 \pm 2.14) ~\mbox{in radian}.
\end{align}
Note that the error in $\delta$ is still very large at this luminosity. 

\begin{figure}[hbtp!]
\centering
\includegraphics[width=0.32\textwidth]{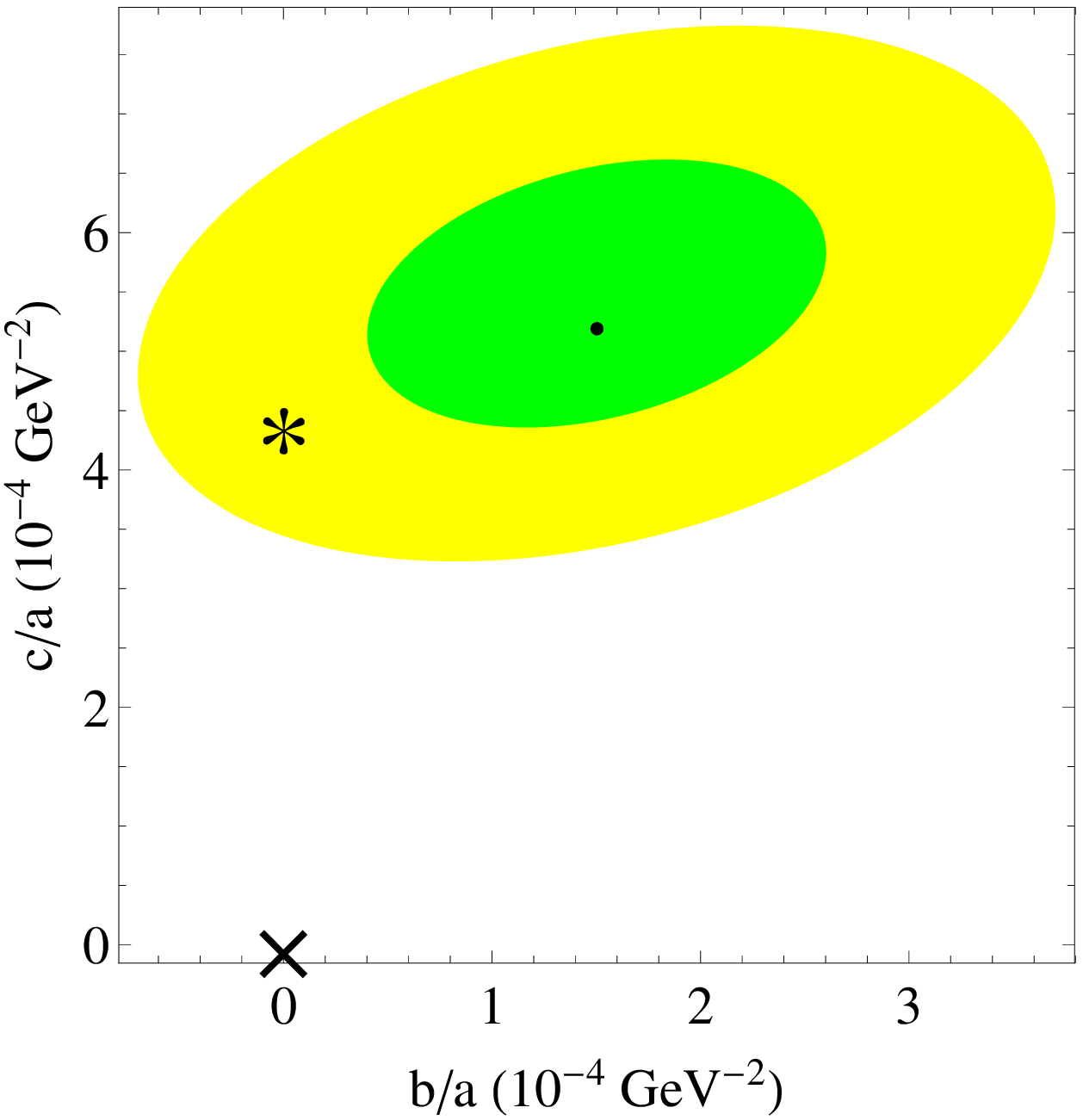} 
\caption{$c/a$ vs $b/a$ $1\sigma$ (green) and $2\sigma$ (yellow)
  contours for CP-odd admixture Higgs at 300 fb$^{-1}$. The best fit value of $(b/a, c/a)$ is shown by the block dot.
  The values with which data are generated $(b/a=0, c/a=4.44 \times 10^{-4})$ is
  shown by the $`*$'. The cross-hair corresponds to $b = c = 0$.}
\label{fig:bacaodd}
\end{figure}

\begin{figure}[hbtp!]
\centering
\includegraphics[width=0.32\textwidth]{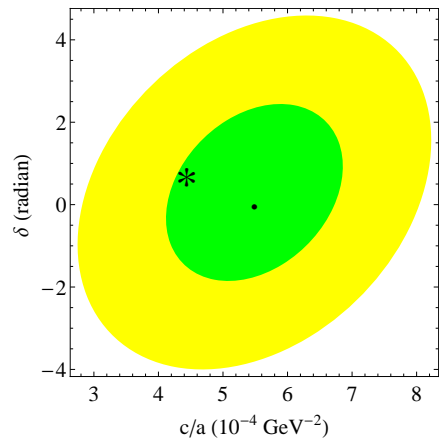} 
\caption{ $\delta$ vs $c/a$  $1\sigma$ (green) and $2\sigma$ (yellow)
  contours for CP-odd admixture Higgs at 300 fb$^{-1}$. The best fit values $(c/a , \delta)$ is
  shown by the block dot. The values with which data are generated is
  shown by the $`*$'.}
\label{fig:dcaodd}
\end{figure}

The best fit values with $1\sigma$ and $2\sigma$ contours for
$c/a$ vs $b/a$ and $ \delta $  vs $c/a$ are shown 
in Fig.~\ref{fig:bacaodd} and  Fig.~\ref{fig:dcaodd} respectively.
\subsection{Study of  angular asymmetries of Higgs at 14 TeV 3000 fb$^{-1}$}
\label{subsec:higgs3000}
High Luminosity LHC (HL-LHC) i.e 14 TeV 3000 fb$^{-1}$ run before 
the energy upgrade will allow us to test CP structure of $HZZ$ couplings even more precisely.
For  3000 fb$^{-1}$ also, we have followed the same cut based analysis that we have discussed
earlier apart from a strong sequential $p_T$ ordering i.e. 
$p_T$  of at least three leptons in a quadruplet must satisfy $p_T >$ 20 GeV.

At 3000 fb$^{-1}$ we revisit the benchmark cases of SM and CP-odd admixture along with two new analysis of CP-even higher
derivative contribution and CP-even-odd scenario.

\subsubsection{SM Higgs and CP-odd admixture Higgs}
First we investigate CP-odd Higgs and SM Higgs and find out the values of angular observables along with their respective errors.
For CP-odd admixture we have again taken $a=0.7$, $b=0$, $c=(2.2+2.2 i)\times10^{-4}$ and SM Higgs $a=1$, $b=0$, $c=0$.
The fit values of the observables $t_2$, $t_1$, $u_2$, $u_1$, $v_2$, $v_1$ for the SM and CP-odd admixture Higgs are tabulated
in Table~\ref{smhiggs-3000} and Table~\ref{oddadmix-3000} respectively.
\begin{table}[hbtp!]
\caption{The values of the observables for the SM Higgs with respective errors at 3000 fb$^{-1}$.}
\centering
\begin{tabular}{c |c }
\hline
\hline
\textbf{Observables} &    \textbf{Values with errors}  \\
\hline

$t_2$ &     $-0.20 \pm 0.04$                         \\

$t_1$  &     $(0.28 \pm 0.35)\times10^{-1}$         \\

$u_2$  &     $ 0.21 \pm 0.19$        \\

$u_1$  &      $(0.46 \pm 2.06 )\times10^{-1}$        \\

$v_2$  &     $(-0.07 \pm 1.96 )\times10^{-1}$        \\

$v_1$  &     $(-0.18 \pm 1.84 )\times10^{-1}$         \\
\hline
\end{tabular}
\label{smhiggs-3000}
\end{table}
The errors have significantly reduced for all the observables and the fit values 
for ratios of couplings for the SM Higgs $b/a$, $c/a$ are given
\begin{align}
b/a &= (0.43 \pm 0.55)\times10^{-4}\,\mbox{GeV}^{-2} \\
c/a &= (1.08 \pm 1.17)\times10^{-4}\,\mbox{GeV}^{-2} 
\end{align}

\begin{figure}[hbtp!]
\centering
\includegraphics[width=0.32\textwidth]{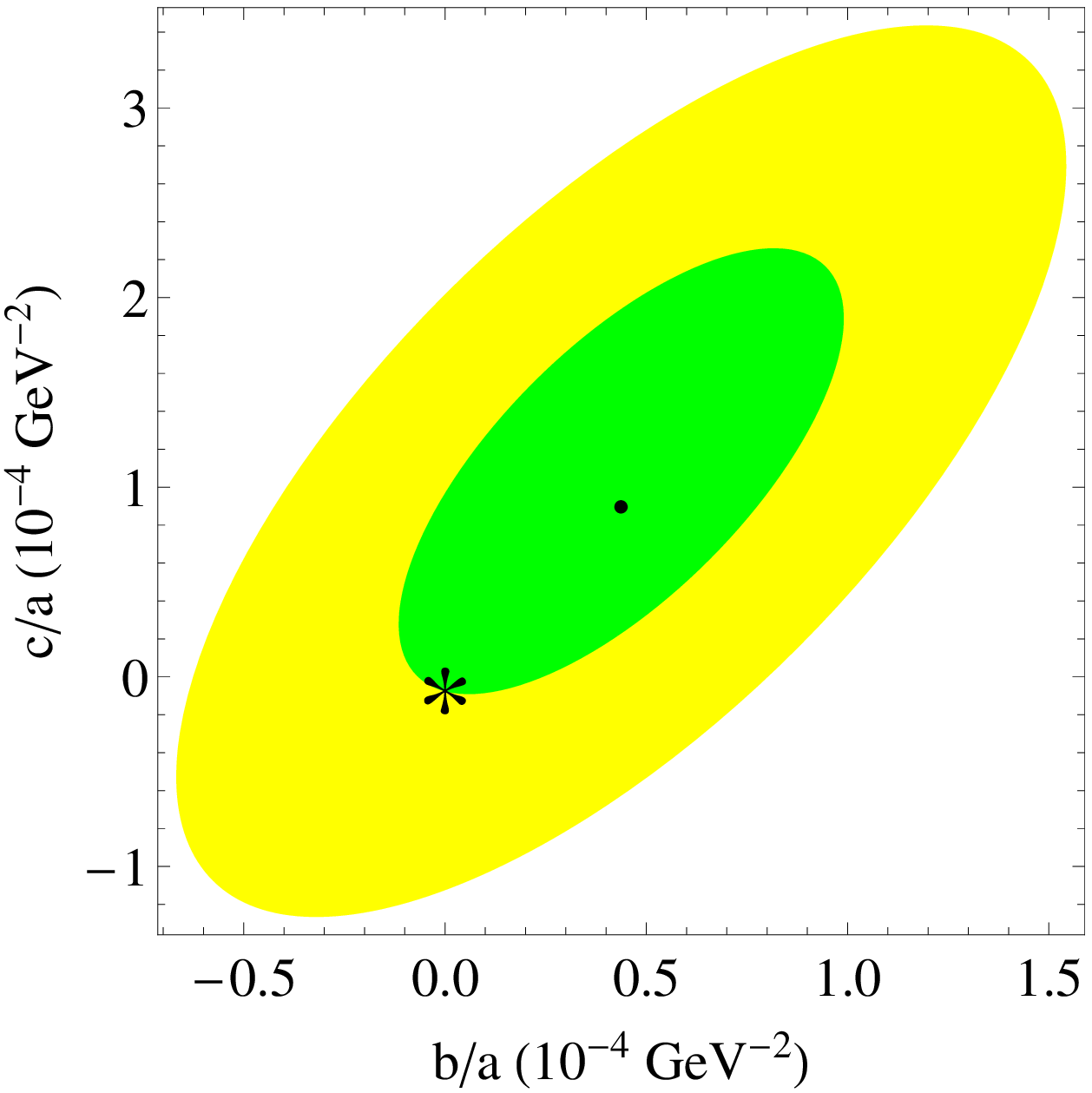} 
\caption{$c/a$ vs $b/a$ $1\sigma$ (green) and $2\sigma$ (yellow)
  contours for the SM Higgs at 3000 fb$^{-1}$.  The best fit values $(b/a, c/a)$ is
  shown by the block dot. The $`*$' corresponds to $b = c = 0$.}
\label{fig:bacasm-3000}
\end{figure}

The 1$\sigma$ and 2$\sigma$ contours for $b/a$ vs $c/a$ are shown in Fig.~\ref{fig:bacasm-3000}

\begin{table}[hbtp!]
\caption{The values of the observables for CP-odd admixture Higgs with respective errors at 3000 fb$^{-1}$}
\centering
\begin{tabular}{c |c }
\hline
\hline
\textbf{Observables} &    \textbf{Values with errors}  \\
\hline

$t_2$ &     $-0.11 \pm 0.04$                         \\

$t_1$  &     $-0.06 \pm 0.03$         \\

$u_2$  &     $0.02 \pm 0.18$        \\

$u_1$  &      $(-0.10\pm 0.56 )\times10^{-1}$        \\

$v_2$  &     $(0.72\pm 1.84 )\times10^{-1}$        \\

$v_1$  &     $(0.67\pm 1.83 )\times10^{-1}$         \\
\hline
\end{tabular}
\label{oddadmix-3000}
\end{table}

At 3000 fb$^{-1}$ from Table~\ref{oddadmix-3000} one can see that the errors in $t_1$ and $t_2$
are much reduced, making them very good observables for probing CP-odd admixture. The best fit values 
for $b/a$, $c/a$ and phase $\delta$  for CP-odd admixture are given as
\begin{align}
b/a &= (0.40 \pm 0.54)\times10^{-4}\,\mbox{GeV}^{-2} \\
c/a &= (3.99 \pm 0.64)\times10^{-4}\,\mbox{GeV}^{-2} \\
 \delta & = 0.45 \pm 1.08 ~\mbox{in radian}.
\end{align}
It should be noted that the error in $\delta$ has become lower due to the fact that the error in $t_1$,
which constraints the phase $\delta$, is much reduced.

\begin{figure}[hbtp!]
\centering
\includegraphics[width=0.32\textwidth]{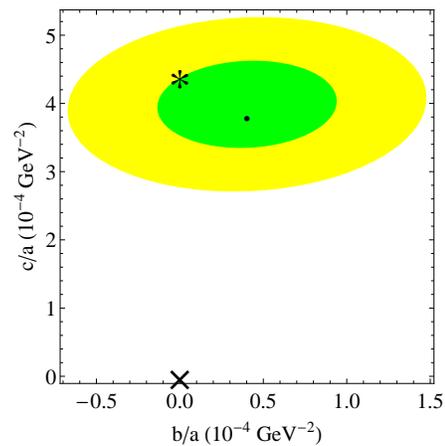} 
\caption{$c/a$ vs $b/a$ $1\sigma$ (green) and $2\sigma$ (yellow)
  contours for CP-odd admixture Higgs at 3000 fb$^{-1}$. The best fit value of $(b/a, c/a)$ is shown by the block dot.
  The values with which data are generated $(b/a=0, c/a=4.44 \times 10^{-4})$ is
  shown by the $`*$'. The cross-hair corresponds to $b = c = 0$.}
\label{fig:bacaodd-3000}
\end{figure}

\begin{figure}[hbtp!]
\centering
\includegraphics[width=0.32\textwidth]{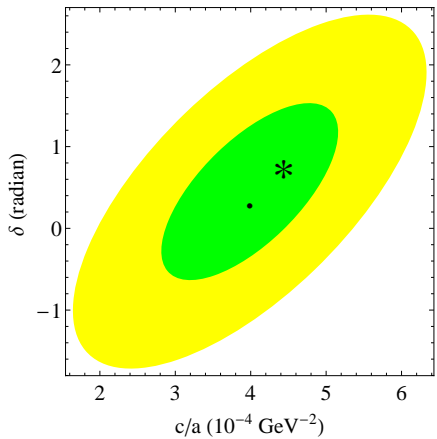} 
\caption{ $\delta$ vs $c/a$  $1\sigma$ (green) and $2\sigma$ (yellow)
  contours for CP-odd admixture Higgs at 3000 fb$^{-1}$. The best fit values $(c/a , \delta)$ is
  shown by the block dot. The cross-hair corresponds to $\delta = c = 0$. 
  The value with which data are generated is shown by the $`*$'.}
\label{fig:dcaodd-3000}
\end{figure}

The 1$\sigma$ and 2$\sigma$ contours for $b/a$ vs $c/a$ and $\delta$ vs $c/a$ are shown in Fig.~\ref{fig:bacaodd-3000}
and Fig.~\ref{fig:dcaodd-3000} respectively.

So far, we have discussed how using angular asymmetries one can probe $HZZ$ couplings
of the SM Higgs and Higgs with mixed CP scenarios at 14 TeV for two different luminosity 300 fb$^{-1}$ and
3000 fb$^{-1}$. The values of the observables vary depending on the values of $a$, $b$ and $c$.
The observables $T_1$, $V_1$ and $V_2$ are sensitive to CP-odd admixture and can be a good candidate to
probe CP-odd admixture. Finally the best fit values for the ratios of the couplings, $b/a$ and $c/a$ are 
calculated using Eq.(\ref{eq:chisq}).

\subsubsection{Higgs with CP-even higher derivative contribution and CP-even-odd scenario}
In this subsection we benchmark the angular asymmetries for the cases:\\
1)Higgs with CP-even higher derivative contribution.  \\
2)Higgs with both CP-odd and CP-even higher derivative contribution i.e. CP-even-odd scenario.\\
For CP-even higher derivative contribution we have taken $a=0.80$, $b=10^{-4}$, $c=0$. However 
for the CP-even-odd scenario we have taken $a=0.75$, $b=8\times10^{-5}$, $c=(1+1i)\times10^{-4}$.

\begin{figure}[hbtp!]
\centering
\includegraphics[width=0.32\textwidth]{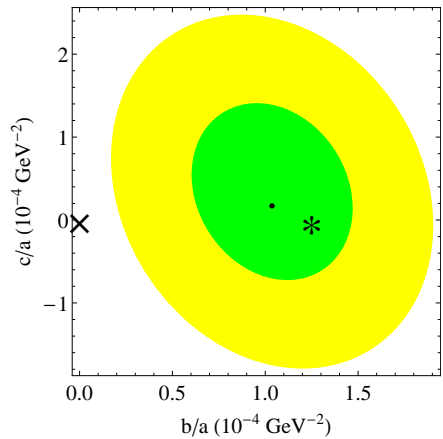} 
\caption{$c/a$ vs $b/a$ $1\sigma$ (green) and $2\sigma$ (yellow)
  contours for Higgs with CP-even higher derivative contribution at 3000 fb$^{-1}$. 
  The best fit value of $(b/a, c/a)$ is shown by the block dot.
  The values with which data are generated $(b/a=1.25 \times 10^{-4}, c/a=0)$ is
  shown by the $`*$'. The cross-hair corresponds to $b = c = 0$.}
\label{fig:bacaeven-3000}
\end{figure}

The values of the observables for CP-even higher derivative contribution are tabulated in Table~\ref{evenadmix-3000}.
\begin{table}[hbtp!]
\caption{The values of the observables for Higgs with CP-even higher derivative 
contribution and their respective errors at 3000 fb$^{-1}$.}
\centering
\begin{tabular}{c |c }
\hline
\hline
\textbf{Observables} &    \textbf{Values with errors}  \\
\hline

$t_2$  &     $-0.12 \pm 0.04$                         \\

$t_1$  &     $(0.31\pm 3.40 )\times10^{-2}$         \\

$u_2$  &     $0.16 \pm 0.18$        \\

$u_1$  &      $(-0.2\pm 1.90 )\times10^{-1}$        \\

$v_2$  &     $(-0.09\pm 1.82 )\times10^{-1}$        \\

$v_1$  &     $(-0.45\pm 1.76 )\times10^{-1}$         \\
\hline
\end{tabular}
\label{evenadmix-3000}
\end{table}
The best fit values for $b/a$ and $c/a$ with errors for CP-even higher derivative contribution are given as
\begin{align}
 b/a &= (1.04 \pm 0.43)\times10^{-4}\,\mbox{GeV}^{-2} \\
 c/a &= (0.34 \pm 1.07)\times10^{-4}\,\mbox{GeV}^{-2} 
\end{align}
The 1$\sigma$ and 2$\sigma$ contours for $b/a$ vs $c/a$ for Higgs with CP-even higher derivative contribution  is shown in Fig.~\ref{fig:bacaeven-3000}.

The values of observables for Higgs in CP-even-odd scenario are tabulated in Table~\ref{evenoddadmix-3000}.
\begin{table}[hbtp!]
\caption{The values of the observables for CP-even-odd scenario with respective errors at 3000 fb$^{-1}$.}
\centering
\begin{tabular}{c |c }
\hline
\hline
\textbf{Observables} &    \textbf{Values with errors}  \\
\hline

$t_2$  &     $-0.13 \pm 0.04$                         \\

$t_1$  &     $0.02\pm 0.03$         \\

$u_2$  &     $0.11 \pm 0.18$        \\

$u_1$  &      $(0.27\pm 1.89)\times10^{-1}$        \\

$v_2$  &     $(0.63\pm 1.82 )\times10^{-1}$        \\

$v_1$  &     $(0.83\pm 1.78 )\times10^{-1}$         \\
\hline
\end{tabular}
\label{evenoddadmix-3000}
\end{table}
It should be noted that value of the observable $t_1$ is large due to non zero CP-odd admixture 
for the CP-even-odd scenario.

At $3000$ fb$^{-1}$ the best fit values of $b/a$, $c/a$, $\delta$ for the CP-even-odd scenario are given as
\begin{align}
 b/a &= (0.59 \pm 0.40)\times10^{-4}\,\mbox{GeV}^{-2} \\
 c/a&= (2.10 \pm 0.87)\times10^{-4}\,\mbox{GeV}^{-2} \\
 \delta & = 0.57 \pm 1.33 ~\mbox{in radian}.
\end{align}

\begin{figure}[hbtp!]
\centering
\includegraphics[width=0.32\textwidth]{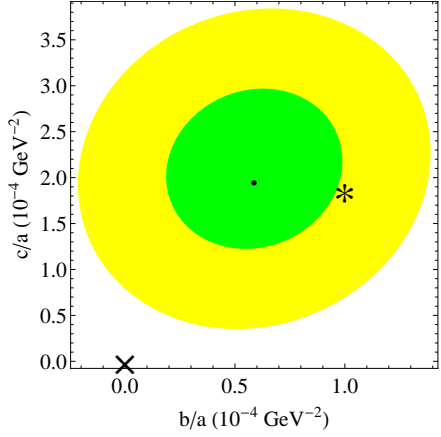} 
\caption{$c/a$ vs $b/a$ $1\sigma$ (green) and $2\sigma$ (yellow)
  contours for CP-even-odd scenario Higgs at 3000 fb$^{-1}$. The best fit value of $(b/a, c/a)$ is shown by the block dot.
  The values with which data are generated $(b/a=1.06 \times 10^{-4},\; c/a=1.89\times 10^{-4})$ is
  shown by the $`*$'. The cross-hair corresponds to $b = c = 0$.}
\label{fig:bacaevenodd-3000}
\end{figure}

\begin{figure}[hbtp!]
\centering
\includegraphics[width=0.32\textwidth]{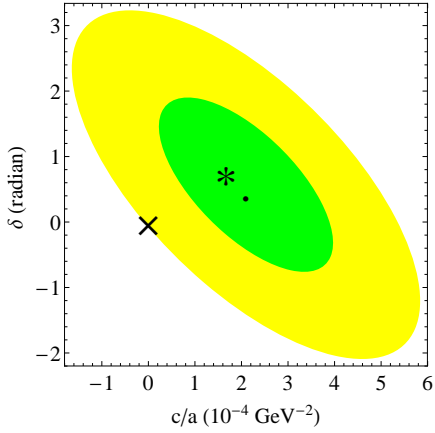} 
\caption{ $\delta$ vs $c/a$  $1\sigma$ (green) and $2\sigma$ (yellow)
  contours for CP-even-odd scenario Higgs at 3000 fb$^{-1}$. The best fit values $(b/a, c/a)$ is
  shown by the block dot. The cross-hair corresponds to $\delta = c = 0$. 
  The value with which data are generated is shown by the $`*$'.}
\label{fig:dcaevenodd-3000}
\end{figure}

The 1$\sigma$ and 2$\sigma$ contours for $b/a$ vs $c/a$ and $\delta$ vs $c/a$ are shown in Fig.~\ref{fig:bacaevenodd-3000}
and Fig.~\ref{fig:dcaevenodd-3000} respectively.


\subsection{Momentum dependence of Higgs couplings}
\label{subsec:madep}
The vertex factors $a$, $b$ and $c$ written in Eq.~(\ref{eq:vertex}) are in general
momentum dependent but in SM they have no momentum dependence. Thus it is essential 
to verify their momentum independence to establish $H$ as the SM Higgs. To achieve this, one has to
measure the momentum dependence of $a$, $b$ and $c$ in different momentum regions.
In LHC one only measures the ratios of couplings i.e. $b/a$ and $c/a$. However one can measure
the values of $b/a$ and $c/a$ in different $q_2^2(m_{34})$ regions.
This will allow us to check the momentum dependence of $b/a$ and $c/a$ and finally
$a$ at 14 TeV $300$ fb$^{-1}$ LHC run.

Integrating Eq.~\eqref{eq:cost1}, \eqref{eq:cost2}, \eqref{eq:phi} 
over $q_2^2$ we get uniangular distribution in $n$-th bin as follows
\begin{align}
\frac{1}{\Gamma^n}\frac{d\Gamma}{d\cos\theta_1} &= \frac{1}{2} -
\mathcal{T}_1^n \, \cos\theta_1 +
\mathcal{T}_2^n \, P_2(\cos\theta_1),\label{eq:Ct1n} \\
\frac{1}{\Gamma^n}\frac{d\Gamma}{d\cos\theta_2} &= \frac{1}{2} +
\mathcal{T}_1^n \, \cos\theta_2 +
\mathcal{T}_2^n \, P_2(\cos\theta_2),\label{eq:Ct2n}\\
\frac{1}{\Gamma^n}\frac{d\Gamma}{d\phi} &= \frac{1}{2\pi} +
\mathcal{U}_1^n \, \cos\phi + \mathcal{U}_2^n
\,\cos2\phi
\nn\\
& \quad  \quad \quad+\mathcal{V}_1^n  \, \sin\phi + \mathcal{V}_2^n 
\,\sin2\phi,\label{eq:Phin}
\end{align} 
where $\mathcal{T}_1^n$, $\mathcal{T}_2^n$, $\mathcal{U}_1^n$, $\mathcal{U}_2^n$
$\mathcal{V}_2^n$ and $\mathcal{V}_1^n$ are observables in $n$-th bin integrated over $q_2^2$ range in that bin.

In SM the vertex factors are momentum independent and have constant
values. This implies that $a$ must have a constant value $1$ and $b$, $c$ 
should be zero in each $m_{34}$ bin. As a result the values $b/a$, $c/a$ will
be zero in each $m_{34}$ bin. In this subsection we bin the events
and find the asymmetries $\mathcal{T}_1^n$, $\mathcal{T}_2^n$,
$\mathcal{U}_1^n$, $\mathcal{U}_2^n$  and utilize them to find out $b/a$ and $c/a$ in 
three different bins. 

\begin{table}[!h]
\caption{Three $\sqrt{q_2^2}=m_{34}$ bins and corresponding number of events in each bins for momentum dependence
measurements at 14 TeV $300$ fb$^{-1}$ LHC run. }
\centering
\begin{tabular}{l|c | c}
\hline \hline
Bin No. &  Bin range $\sqrt{q_2^2}=m_{34}$               & Number of events  \\
\hline
Bin 1 & $12.00\mbox{ GeV} < m_{34} < 29.00$~GeV       &  $210$             \\
Bin 2 &$29.00\mbox{ GeV} < m_{34} < 46.00$~GeV        &  $187$               \\
Bin 3 & $46.00\mbox{ GeV} < m_{34} < 80.00$~GeV       &  $52$               \\
\hline
\end{tabular}
\label{bin_table_1}
\end{table}

Now we will study the momentum dependence of vertex factor `$a$' in these 3 bins,
we rewrite the decay width in $n$-th bin $\Gamma_n$ as follows
\begin{equation}
 \Gamma_n=a_n^2\Gamma'_n\left( b/a, c/a \right)\label{bbya}.
\end{equation}
where $a_n$ is the value of $a$ in $n$-th bin. $\Gamma'_n$ is obtained by dividing $\Gamma_n$ by $a^2$ and 
making it a function of $b/a$ and $c/a$ only. We can calculate the values of $\Gamma'_n$ and its errors in 
different bins by substituting the values of $b/a$ and $c/a$ from Table~\ref{baca}.

\begin{table}[!h]
\centering
\caption{$b/a$ and $c/a$ with corresponding errors in 3 bins in $\mbox{GeV}^{-2}$}
\begin{tabular}{|l|c| c|  }
\hline
Bin No. & $b/a$  in $10^{-4} \mbox{GeV}^{-2}$  &   $c/a$ in $10^{-4} \mbox{GeV}^{-2}$\\     
\hline
Bin 1 &    $0.02\pm 0.44$            &     $0.30\pm0.57$                        \\ 
Bin 2 &    $0.39\pm 0.79$            &     $0.60\pm0.89$                        \\
Bin 3 &    $1.35\pm 2.09$            &     $1.56\pm2.24$                                      \\
\hline        
\end{tabular}
\label{baca}
\end{table}
Fit values of $b/a$ and $c/a$ have relatively larger errors in last bin due to small number
of events and both $b/a$ and $c/a$ are consistent with zero in each $m_{34}$ bins.

For resonant production of Higgs we can factor out the production cross-section 
and  $\Gamma_n$ is proportional to the number of events $N_n$ 
in each of the three $m_{34}$ bins.  If $\Gamma_i$, $\Gamma_j$ and $N_i$ , $N_j$
are the value of decay widths and number of events in $i$ and $j$-th bin respectively,
then
\begin{equation}
 \frac{\Gamma_i}{\Gamma_j} = \frac{N_i}{N_j} \label{gamr}
\end{equation}
will hold between any two bins.
Furthermore putting $\Gamma_i$ and $\Gamma_j$ expressions as written in Eq.~\eqref{bbya}
into Eq.\eqref{gamr} one finds the following relationship
\begin{equation}
   r_{ij}= \frac{a_i}{a_j}=\sqrt{ \frac{N_i}{N_j} \frac{\Gamma'_j}{\Gamma'_i}} \label{ratio}
\end{equation}
If $a$ is independent of momentum the ratio $r_{ij}$ will always be $1$ for any two bins.
We tabulate all possible $r_{ij}$ in Table.~\ref{a_rtio} for this three bins with their corresponding
errors. One can similarly perform the same analysis at 3000 fb$^{-1}$. At 3000 fb$^{-1}$ the errors
in $b/a$ and $c/a$ will be reduced and with the enhanced statistics, one may in principle have enough
events to increase the number of bins to check momentum dependence of Higgs couplings. 
\begin{table}[!h]
\centering
\caption{Ratio $r_{ij}$ between different bins.}
\begin{tabular}{c |c  }
\hline
\hline 
Ratios ($r_{ij}$)                 & values with errors\\
\hline
$r_{12}$ \hspace{.8cm}  &  $1.06\pm0.28$ \\ 
$r_{13}$\hspace{.8cm}   &  $0.82\pm0.38$ \\
$r_{23}$\hspace{.8cm}   &  $0.75\pm 0.39$  \\
\hline        
\end{tabular}
\label{a_rtio}
\end{table}

These Ratios are consistent with SM (i.e ratio  $r_{ij} =1$) as they should. The ratio $r_{13}$ and $r_{23}$
have larger errors due to small statistics. These results will be improved by the bin size optimization and 
with larger statistics.

\section{Comparison of precision measurement between 33 TeV LHC and ILC}
\label{comp}
After HL-LHC (High Luminosity LHC), LHC energy could be upgraded to run at a center of mass energy of $\sqrt{s}=33$ TeV.
In this section we compare the precision measurements of $H$ coupling between $33$ TeV LHC and ILC. We
have followed the same cut based analysis as 14 TeV 3000 $fb^{-1}$ machine.
At 33 TeV 3000 fb$^{-1}$ the number of events are given in the cut flow Table~\ref{cut33}.  
\begin{table}[hbtp!]
\centering
\begin{tabular}{|c|c|}
\hline
Cuts & Number of events  \\
\hline
Selection  cuts                           & 6523      \\
$50\mbox{ GeV} <m_{12} < 106$~GeV         & 6362      \\
$12\mbox{ GeV} <m_{34} < 115$~GeV         & 5935      \\
$115 \mbox{ GeV} < m_{4\ell} < 130$~GeV   & 5852      \\
\hline
\end{tabular}
\caption{Effect of the sequential cuts on the simulated Signal for $33$ TeV $3000$ fb$^{-1}$ LHC upgrade.
The $k$-factor for signal is $1.95$.}
\label{cut33}
\end{table}
The best fit values of $b/a$ and $c/a$ for SM Higgs are  
\begin{align}
 b/a &= (2.64 \pm 1.60)\times10^{-5}\,\mbox{GeV}^{-2} \\
 c/a&= (1.07 \pm 6.25)\times10^{-5}\,\mbox{GeV}^{-2} 
\end{align}

\begin{figure}[hbtp!]
\centering
\includegraphics[width=0.32\textwidth]{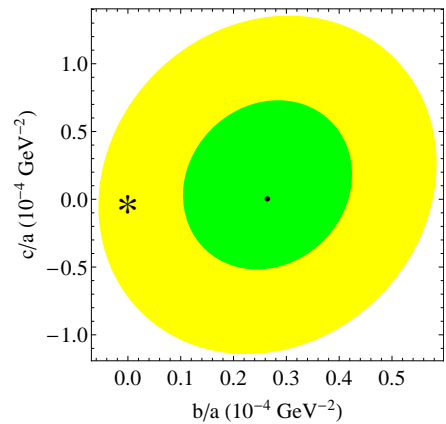} 
\caption{$c/a$ vs $b/a$ $1\sigma$ (green) and $2\sigma$ (yellow)
  contours for SM Higgs at 33 TeV 3000 fb$^{-1}$ LHC run.  The best fit values $(b/a, c/a)$ is
  shown by the block dot. The $`*$' corresponds to $b = c = 0$.}
\label{fig:bacasm33TeV}
\end{figure}

In our first attempt, we find that a precision  of $~10^{-5}$ is achievable at the $33$ TeV  3000 fb$^{-1}$ LHC,
for CP-odd admixture and CP-even higher derivative contribution.
In future, a more detailed analysis may improve the result and decrease of errors in $b/a$ and $c/a$. 
Although the precision achieved is high, LHC will only measure the ratios of Higgs couplings even at $33$ TeV. 

ILC will provide an independent test in measuring $HZZ$ couplings
subject to its proposed implementation and will also offer invaluable probe to such couplings.
At LHC one measures $\sigma\cdot BR(H \rightarrow ZZ)$ however ILC will measure the
branching ratio $BR(H \rightarrow ZZ)$ by measuring the inclusive cross section
$\sigma_{ZH}$ for the process $e^+ e^-\rightarrow Z H$. This inclusive measurement of cross section
alone will probe $HZZ$ couplings to $1.3 \%$\cite{Baer:2013cma}. Identifying a 
$Z$ boson in recoil against the Higgs boson one can find out the partial 
width $\Gamma(H \rightarrow ZZ)$
\begin{align}
\Gamma_{total}=\frac{\Gamma(H \rightarrow ZZ)}{BR(H \rightarrow ZZ)}.
\end{align}

For example the expected precision for CP-odd anomalous coupling $c$ at ILC\cite{Asner:2013psa} 
is $7\times 10^{−4}$ to $8\times 10^{−6}$ which is roughly around the loop induced CP-odd contribution. 
We have shown that precision in the measurement of $c/a$ can be $6\times10^{-5}$ for $33$ TeV LHC
using angular asymmetries. However at ILC the measurement of Higgs decay width
will allow us to extract the absolute values $a$, $b$, $c$ which is beyond
the scope of LHC.
\section{Conclusion}
\label{sec:conclusion}
We demonstrate that angular asymmetries will provide a strong and efficient tool to probe Higgs couplings
in high luminosity future LHC runs. With the increased statistics at 14 TeV run, LHC will enter into precision
era and angular analysis will offer a step by step methodology to study the Higgs couplings. Angular 
asymmetries can be utilized to probe the Higgs couplings and to disentangle its exact CP property 
in the next LHC run. We benchmark our observables for SM, CP-odd admixture, CP-even higher derivative contribution
and finally CP-odd admixture with Higher derivative CP-even coupling. We perform the analysis for two 
different luminosities at 300 fb$^{-1}$ as well as 3000 fb$^{-1}$ and study the precision with which angular analysis 
probe $HZZ$ couplings. The study of the momentum dependence of the Higgs couplings would be a significant step 
forward in establishing its SM nature, since in the SM Higgs couplings do not have any momentum dependence.
At 14 TeV LHC run with the improved statistics, we present how one can examine the 
momentum dependence of the Higgs couplings in different momentum regions. We have further 
discussed what precision LHC can achieve in the measurement of the Higgs couplings to Z boson by angular analysis
at 33 TeV. A comparison has also been made between precision measurement of 33 TeV LHC and ILC. 
Angular analysis will be a powerful technique to decipher the CP properties of Higgs couplings at 14 TeV LHC run
and will open up a new domain of precision measurement. 

\acknowledgments
Work of BB is supported by Department of Science and Technology, Government of INDIA under the Grant Agreement numbers IFA13-PH-75 (INSPIRE Faculty Award).

\appendix
\section{}
The amplitudes for the process $H$ with spin $\mathbf{J}(spin\;0)$ decays to  two $Z(spin\; 1)$
boson with spin projections along $z$ axis $\lambda_1$ and $\lambda_2$ is\cite{Jacob:1959at,Chung:1102240} 
\begin{align}
  \mathscr{M}(J_z, \lambda_1, \lambda_2) &= \left( \frac{2J + 1}{4\pi}
  \right)^{\frac{1}{2}} \mathscr{D}_{J_z
    \lambda}^{J*}(\Phi,\Theta,-\Phi) \; \mathcal{A}_{\lambda_1, \lambda_2},
\end{align}
where $\mathscr{D}_{J_z\lambda}^{J*}$ is the Wigner-$D$ function, $\lambda=\modulus{\lambda_1-\lambda_2}$ and  
$\mathcal{A}_{\lambda_1 ,\lambda_2}$ are the \textit{helicity amplitudes} with $\lambda_{1,2}
\in\{\pm 1,0\}$, $J = \modulus{\mathbf{J}}$.
Angular momentum conservation implies $\modulus{\lambda} = \modulus{\lambda_1 - \lambda_2} \leqslant J$
and the helicity amplitudes are related as $\mathcal{A}_{\lambda_2,\lambda_1} = (-1)^J \mathcal{A}_{-\lambda_1 ,-\lambda_2}$.
We have three orthogonal helicity amplitudes $\mathcal{A}_{00}$, $\mathcal{A}_{++}$ and $\mathcal{A}_{--}$. However 
one can write three helicity amplitudes in transversity basis as:
\begin{align}
\mathcal{A}_L &= \mathcal{A}_{00} \\
\mathcal{A}_{\parallel}&=\frac{1}{\sqrt{2}}(\mathcal{A}_{++}+\mathcal{A}_{--}) \\
\mathcal{A}_{\perp}&=\frac{1}{\sqrt{2}}(\mathcal{A}_{++}+\mathcal{A}_{--}).
\end{align}

Helicity fractions $\mathcal{F}_L$, $\mathcal{F}_{\parallel}$ and
$\mathcal{F}_{\perp}$ are defined as $\dsp \mathcal{F}_\lambda=\frac{\mathcal{A}_\lambda}{\sqrt{ \modulus{\mathcal{A}_L}^2 +
      \modulus{\mathcal{A}_{\parallel}}^2 + \modulus{\mathcal{A}_{\perp}}^2}}$, where $\lambda \in \{ L, \parallel, \perp \}$.\\

Moreover $\eta=\frac{2 v_{\sss\ell} a_{\sss\ell}}{v_{\sss\ell}^2+a_{\sss\ell}^2}$ with 
$v_{\sss\ell}=2 I_{3\ell}-4e_{\sss\ell} \sin^2\theta_W$ and $a_{\sss\ell}=2 I_{3\ell}$.

$\gammaf$ is defined as 
\begin{equation}
\gammaf = \frac{d \Gamma}{dq_2^2} = \mathcal{N} \left(\modulus{\mathcal{A}_L}^2 + \modulus{\mathcal{A}_{\parallel}}^2 +
    \modulus{\mathcal{A}_{\perp}}^2 \right) \label{eq:gamma}, 
\end{equation}
with  $\mathcal{N} = \frac{1}{2^{4}} \frac{1}{\pi^2} 
  \frac{g^2}{\cos^2{\theta_W}} \frac{\text{Br}_{\ell\ell}^2}{M_H^2} \frac{\Gamma_Z}{M_Z}\frac{X}{\left(\left( q_2^2 - M_Z^2 \right)^2 +M_Z^2 \Gamma_Z^2 \right)}$. $\Gamma_Z$ is the decay width of $Z$,
$\text{Br}_{\ell\ell}$ is the branching fraction for the decay of $Z$ to two massless leptons.
However we assumed narrow width approximation for on-shell $Z_1$ boson in Sec.\ref{sec:form}. In Sec.\ref{sec:num} 
we have implemented the the cut flow table while integrating over $q_1^2$ and $q_2^2$ to find the expressions for
$\mathcal{T}_1$, $\mathcal{T}_2$, $\mathcal{U}_1$, $\mathcal{U}_2$, $\mathcal{V}_1$ and $\mathcal{V}_2$.

The expressions for $\mathcal{T}_1$, $\mathcal{T}_2$, $\mathcal{U}_1$, $\mathcal{U}_2$, $\mathcal{V}_1$ and $\mathcal{V}_2$ 
are 
\begin{widetext}
 \begin{align}
\mathcal{T}_1 &=\frac{1.32\times10^{-9}y}{5.57\times10^{-8}+2.61\times10^{-8}x +3.98\times10^{-9}x^2 + 1.60\times10^{-10}y^2}\\
 \mathcal{T}_2
&=\frac{-9.65\times10^{-9}+4.00\times10^{-10}x^2 + 4.00\times10^{-10}y^2}{5.57\times10^{-8}+2.61\times10^{-8}x +3.98\times10^{-9}x^2 + 1.60\times10^{-10}y^2}\\
\mathcal{U}_1 &=\frac{-1.20\times10^{-9}-6.33\times10^{-10}x + 7.11\times10^{-11}x^2}{5.57\times10^{-8}+2.61\times10^{-8}x +3.98\times10^{-9}x^2 + 1.60\times10^{-10}y^2}\\
 \mathcal{U}_2 &=\frac{6.06\times10^{-9}+4.35\times10^{-9}x+7.97\times10^{-10}x^2 + 4.00\times10^{-10}y^2}{5.57\times10^{-8}+2.61\times10^{-8}x +3.98\times10^{-9}x^2 + 1.60\times10^{-10}y^2}\\
\mathcal{V}_1 &=\frac{3.11\times10^{-10}y}{5.57\times10^{-8}+2.61\times10^{-8}x +3.98\times10^{-9}x^2 + 1.60\times10^{-10}y^2}\\
\mathcal{V}_2 &=\frac{-2.92\times10^{-9}y}{5.57\times10^{-8}+2.61\times10^{-8}x +3.98\times10^{-9}x^2 + 1.60\times10^{-10}y^2}
\end{align}
where $x = \frac{b}{a}\times(100\mbox{Gev})^2$  and $y = \frac{c}{a}\times(100\mbox{Gev})^2$
\end{widetext}

\end{document}